\documentclass[preprint,review,12pt]{elsarticle}
\usepackage{amsmath}
\usepackage{graphicx}
\usepackage{amssymb}
\usepackage{color,soul}
\usepackage[T1]{fontenc}
\usepackage{ae,aecompl}

\begin{document}

\title{Manipulating graphene kinks through positive and negative radiation pressure effects}
\author[niic,nsu]{R.~D.~Yamaletdinov}
\author[ipju]{T.~Roma\'nczukiewicz}
\author[usc]{Y.~V.~Pershin\corref{cor1}}

\cortext[cor1]{Corresponding author. Tel: +1 803 777-5073. Email: pershin@physics.sc.edu (Yuriy Pershin)}

\address[niic]{Nikolaev Institute of Inorganic Chemistry SB RAS, Novosibirsk 630090, Russia}
\address[nsu]{Novosibirsk State University, Novosibirsk 630090, Russia}
\address[ipju]{Institute of Physics, Jagiellonian University, Krak\'ow, Poland}
\address[usc]{Department of Physics and Astronomy, University of South Carolina, Columbia, South Carolina 29208, USA}



\begin{abstract}
We introduce an idea of experimental verification of the counterintuitive negative radiation pressure effect in some classical field theories
by means of buckled graphene.  In this effect, a monochromatic plane wave interacting with topological solutions  pulls these solutions towards the source of radiation.
Using extensive molecular dynamics simulations, we investigate the traveling wave-induced motion of kinks in buckled graphene nanoribbons.
It is shown that depending on the driving source frequency, amplitude and direction, the kink behavior varies from attraction to repulsion (the negative and positive
radiation pressure effects, respectively). Some preliminary explanations are proposed based on the analogy to certain field theory models.
Our findings open the way to a new approach to motion control on the nanoscale.
\end{abstract}
\maketitle   

\section{Introduction}

In the classical mechanics an object colliding with another object at rest transfers its momentum to the latter
making it move (in the simplest 1D case) in the direction of the initial velocity. Similarly,
the electromagnetic radiation hitting a surface exerts the pressure upon the surface due to the exchange of momentum
(the most spectacular example of this phenomenon is the Kepler's observation that the tail of a comet always points away from the Sun).
 Also, in classical field theory models \cite{Tomasz08a}, the radiation source normally exerts an outward pressure on  topological solutions. Using the standard linearized single-channel scattering theory one can show that, typically, the radiation pressure is $ \sim A^2 |R|^2$, where $A$ is the radiation amplitude and $R$ is the reflection coefficient. The classical $\phi^4$ model~\cite{rajaraman1982instantons}, however, is a notable exception characterized by $R\equiv 0$ (see Ref. \cite{Tomasz08a}). In the linear approximation, its topological solutions -- kinks and antikinks -- experience the radiation pressure defined by higher-order corrections leading to a power law of the form $F \sim -A^4 $. Such a counterintuitive behaviour when the force is directed towards the radiation source has been dubbed as the {\it negative radiation pressure} (NRP)
effect~\cite{Tomasz08a}.

Technically, the NRP effect in pure $\phi^4$ model can be explained as follows. In the first (linear)
order in the amplitude of incoming wave,
the kinks are transparent to the radiation. In the second order of the perturbation series, the nonlinearity is the source of waves having the frequency  twice the frequency of incident radiation. These newly created waves carry more momentum than the original waves. To balance the momentum surplus created behind a kink, the kink accelerates towards the radiation source. Another possible NRP mechanism is the multichannel scattering leading to the momentum surplus already in the linear approximation~\cite{Romanczukiewicz:2008hi, Forgacs:2013oda}.

In this letter, we propose a possible experimental verification of the NRP effect based on a buckled graphene nanoribbon~\cite{Mashoff10a,Lindahl12a} or similar physical system.
Recently, it was shown~\cite{Yamaletdinov17b} that the graphene buckled over a trench in a
nonstandard geometry (such that the trench length is much longer than its width)
can reveal some of the properties of
$\phi^4$ model. These include the existence of  kinks and antikinks, as well as their nontrivial dynamics leading, e.g., to annihilation of
topological solutions in low-velocity collisions, and their reflection when they collide at higher speeds. Below, we further explore this analogy showing that under appropriate driving conditions the
graphene kinks experience the negative radiation pressure. Our key observation is that the response of graphene kinks  is defined by a combination of
repulsive and attractive forces that, in the framework of a perturbed model, can be represented by the fourth-order polynomial
\cite{Tomasz08a}
\begin{equation}
F \sim A^2 |R|^2-\mathcal{O}(A^4)\; .
\label{F_phi4}
\end{equation}
Using molecular dynamics simulations we explicitly demonstrate that depending on the driving source frequency, amplitude and direction, the kink behavior varies from attraction to repulsion, in agreement with the general form of Eq. (\ref{F_phi4}). These results may find application in the motion control at the nanoscale.

The analysis of NRP effect in graphene, however, is a non-trivial task. For instance, contrary to the standard $\phi^4$ model, graphene sheets support waves with several possible polarizations, each type of waves having different
dynamical properties. In the multichannel scattering process, the force $F$ exerted on a kink calculated even in the linear approximation may lead to the negative radiation pressure. Assuming that the incoming radiation is predominantly contained in the $i$-th channel, the force $F$ can be written as
\begin{equation}
 F\sim \mathcal{P}_i+\sum_{j}\frac{k_j}{k_i}\left(|R_{ij}|^2-|T_{ij}|^2\right)\mathcal{P}_{i},
 \label{eq:force}
\end{equation}
where $\mathcal{P}_i$ is the momentum flux of the $i$-th component of radiation (usually proportional to $A_i^2 k_i$), $k_{i(j)}$ is the wave number, and $|R_{ij}|^2$ and $|T_{ij}|^2$ are scattering probabilities from $i$-th channel into the forward- and back-propagating states of $j$-th channel \cite{Forgacs:2013oda}. It follows from Eq. (\ref{eq:force}) that a favorable condition for NRP is when all the scattering probabilities into reflected states, $|R_{ij}|^2$, are small and one of the transmission probabilities $|T_{ij}|^2$ into a certain high-momentum channel $j$ is large.

\section{System setup and simulation details}

\begin{figure}[t]%
\centering \includegraphics[width=100mm]{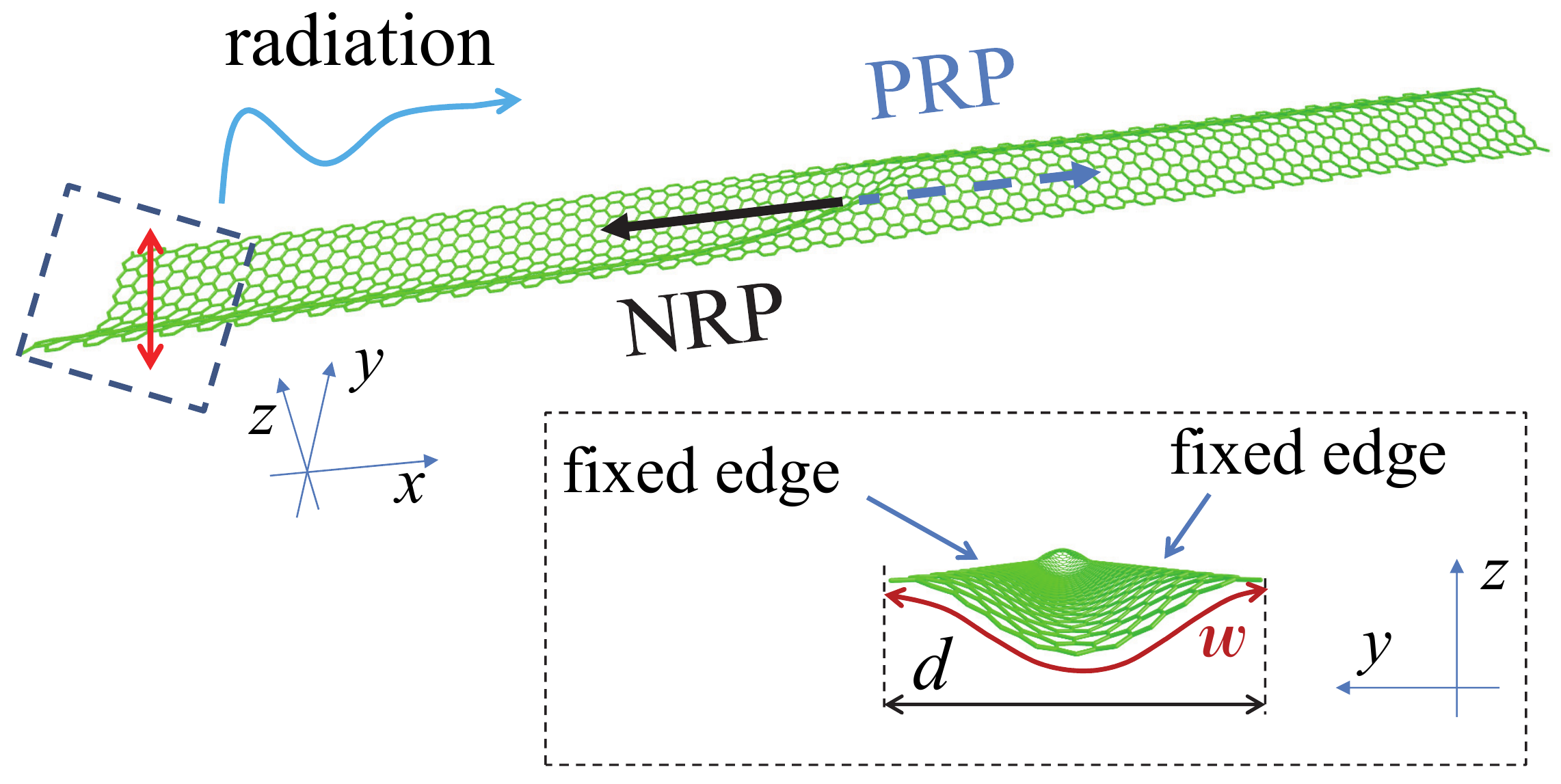}
\caption{Numerical experiment setup. The radiation (created by a sinusoidal force applied to a group of atoms in the dashed rectangle) propagates along the nanoribbon and interacts with an  initially immobile  kink (in the middle). Inset: side view of the setup. The nanoribbon is buckled by changing the distance between the longer edges from $w$ (the free length of the nanoribbon in $y$-direction) to $d<w$. Two first lines of atoms at each longer edge are kept fixed.}
\label{fig1}
\end{figure}

Fig. \ref{fig1} presents the numerical experiment setup.
The dynamics of a buckled graphene nanoribbon with clamped boundary conditions for the longer armchair
edges and free boundary conditions for the shorter edges
 was simulated. The nanoribbon was buckled by changing the
distance between the fixed edges from $w$ to $d < w$. The clamped boundary conditions were implemented by fixing two first lines of carbon atoms of
longer edges. The carbon atoms at shorter edges were saturated by hydrogen atoms.
Our main calculations (Figs. \ref{fig2} and \ref{fig3}) were performed using a nanoribbon
of $L = 214$~nm length and $w = 22$~\r{A} width. Figs. \ref{fig4}, \ref{fig5} were obtained using a shorter nanoribbon of $L = 42$~nm length and $w = 22$~\r{A} width without saturation by hydrogen. The results for the shorter and longer nanoribbons are in a very good agreement.

As the initial condition, we used a nanoribbon with an immobile kink located at the distance of 21 nm from the left edge. The radiation (phonons) was created by applying a sinusoidal force in $x$-, $y$- or $z$-direction to a group of atoms near a shorter (left) edge of the nanoribbon (the atoms located within 23 \r{A} from the edge), unless otherwise stated. In our calculations, these atoms played the role of a radiation source.   This Letter reports the results obtained for $d/w = 0.9$ and $T=0$. We emphasize that in the present context, the radiation is the nanoribbon vibrations induced near the left edge of the nanoribbon (see Fig. \ref{fig1}).

Molecular dynamics simulations were conducted using {NAMD2} software package~\cite{phillips05} ({NAMD} was developed by the Theoretical and Computational Biophysics Group in the Beckman Institute for Advanced Science and Technology at the University of Illinois at Urbana-Champaign).
The interactions between the carbon atoms were described using the standard 2-body spring bond, 3-body angular bond (including the Urey-Bradley term), 4-body torsion angle and Lennard-Jones potential energy terms~\cite{JCC21367}.
We used a previously developed set of force-field parameters for carbon atoms~\cite{Yamaletdinov17b} that
matches the experimentally observed \cite{Lambin14} in-plane stiffness ($E_{2D}=342$~N/m), bending rigidity ($D=1.6$ eV) and equilibrium bond length ($a=1.421$ \r{A}) of graphene. The Lennard-Jones potential's parameters were fitted to match the interlayer distance and adhesion energy in graphites~\cite{Chen2013}. MD simulations were performed with 1 fs time step.
The van der Waals interactions were gradually cut off starting
at 10 \r{A} from the atom until reaching zero interaction 12~\r{A} away.

\section{Results}

\begin{figure}[htb]
\begin{center}
(a) \includegraphics[width=0.44\textwidth]{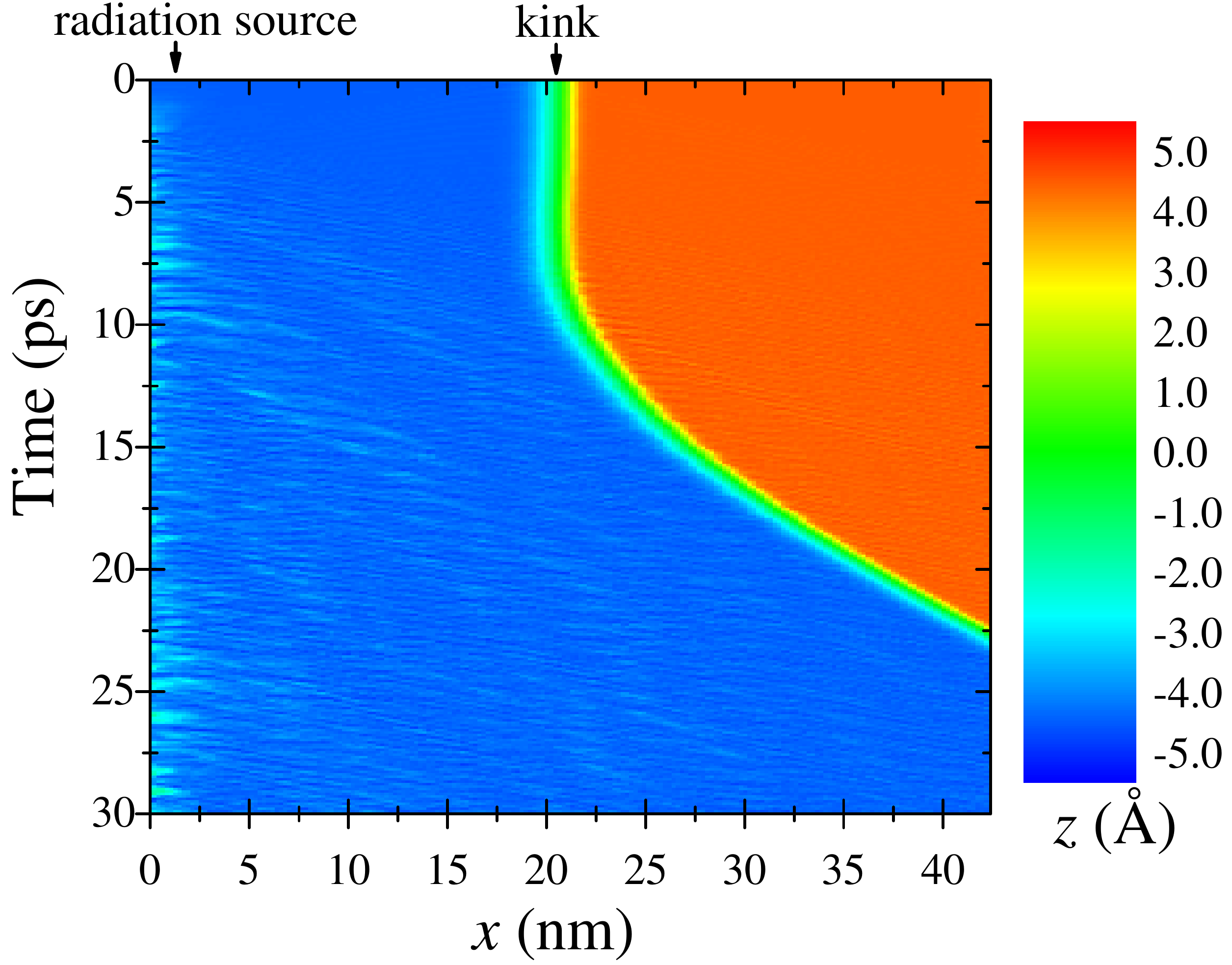}
\qquad
\includegraphics[width=0.44\textwidth]{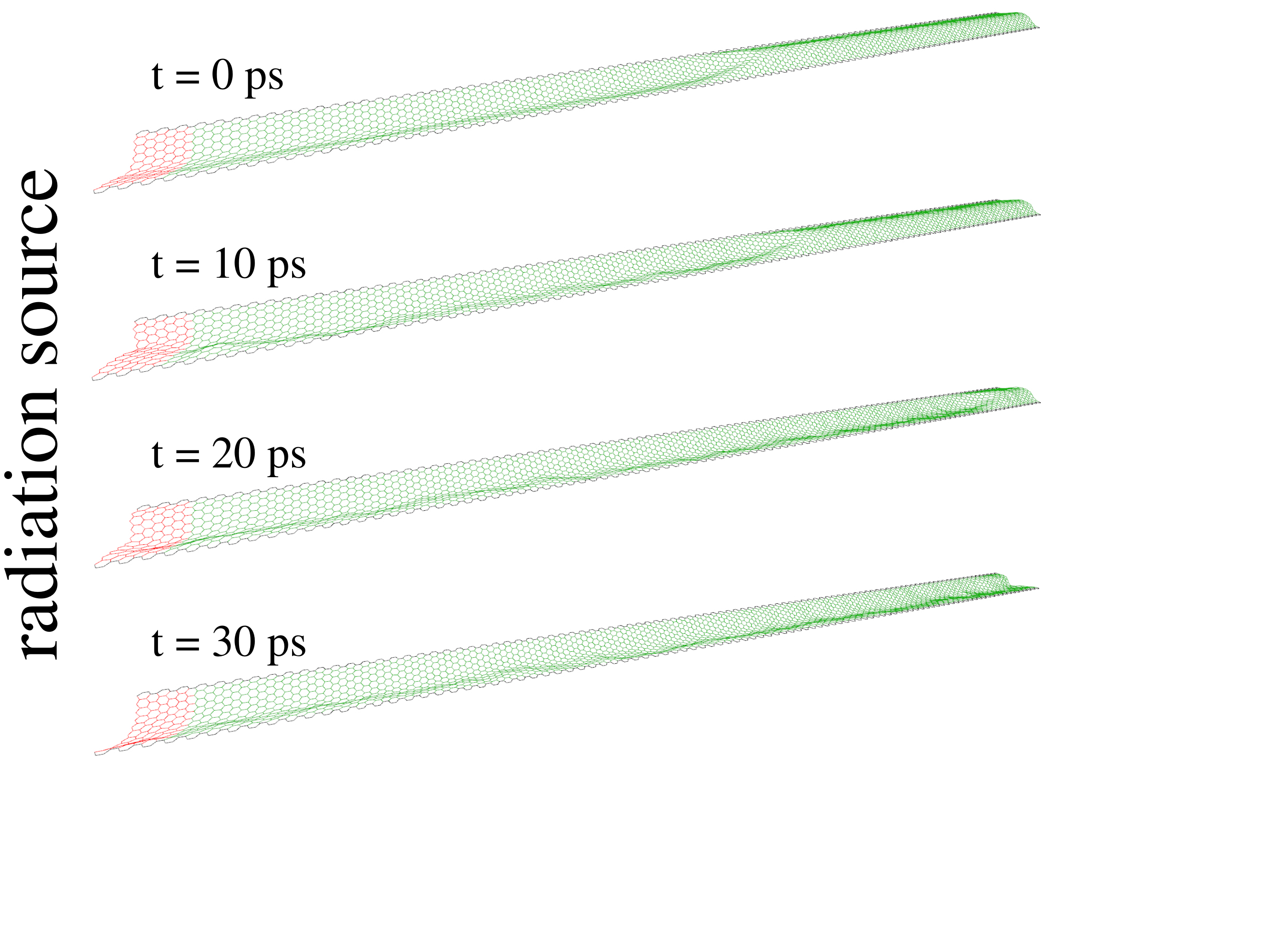}
\\
\vspace{2mm}
(b)  \includegraphics[width=0.44\textwidth]{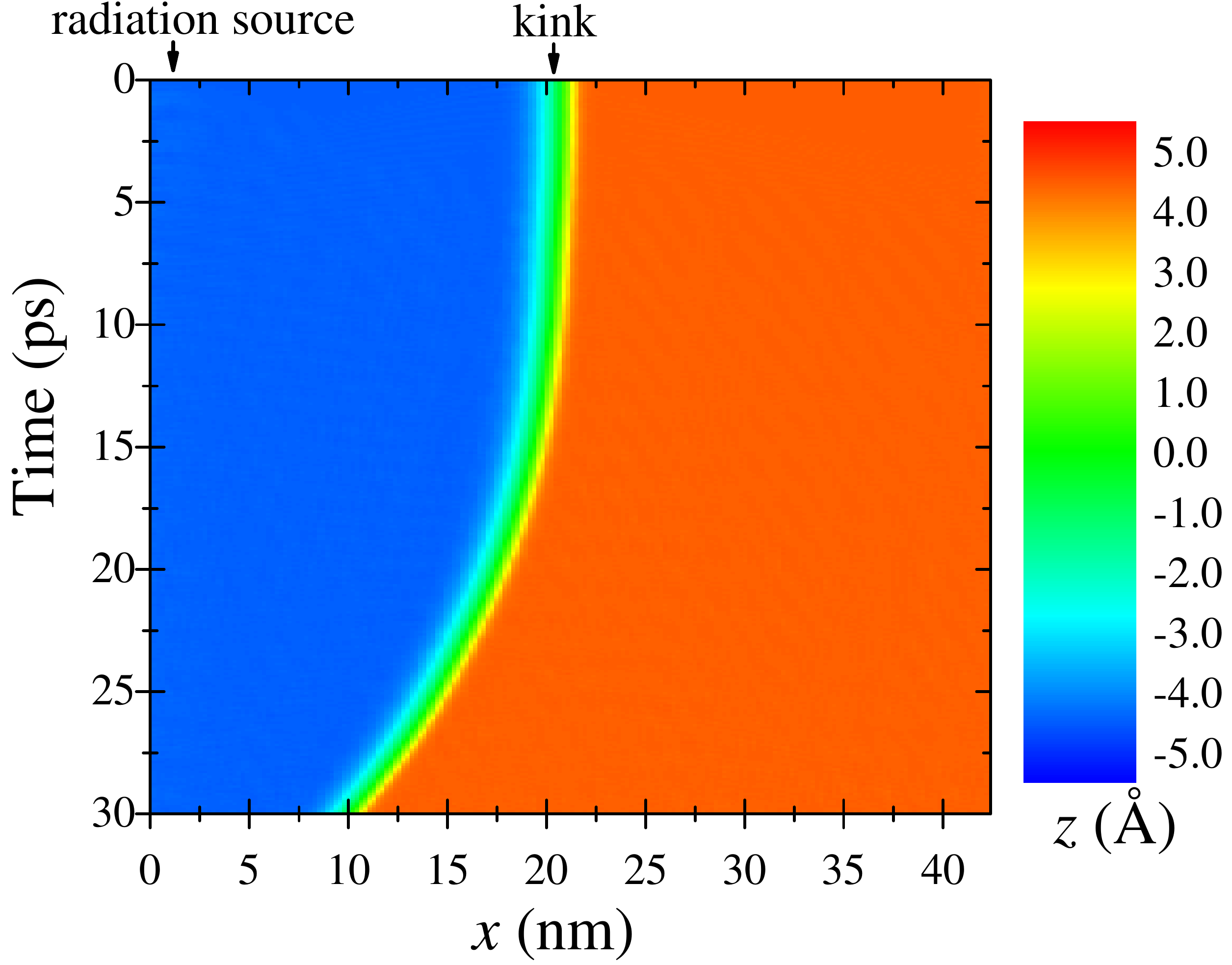}
\qquad
\includegraphics[width=0.44\textwidth]{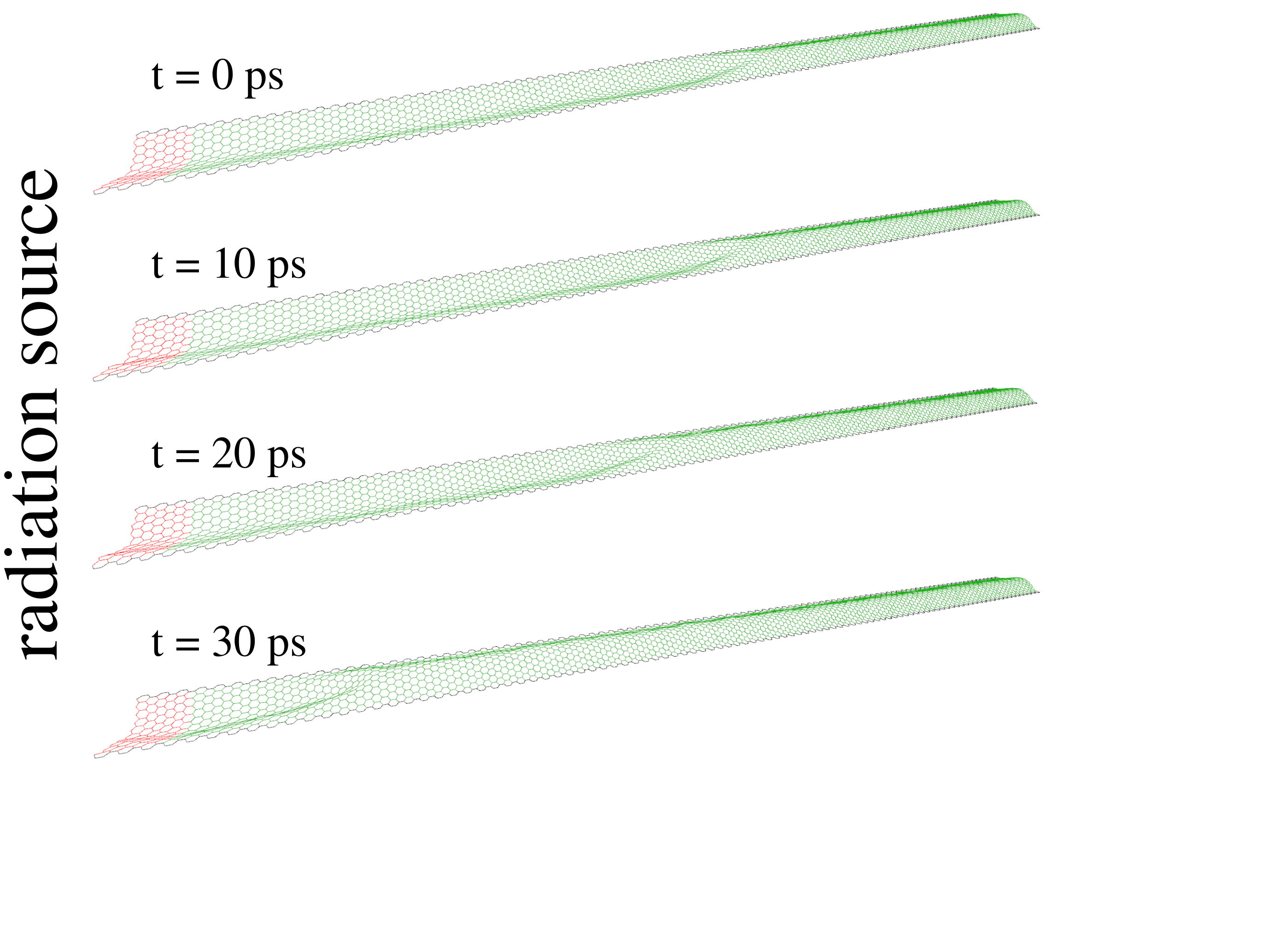}
\end{center}
\caption{Positive (a), and negative (b) radiation pressure effects. Here, the color represents $z$-coordinates of the central chain of atoms along the membrane (the atoms with $y$-coordinate close to zero). The position of kink corresponds to the green color. The radiation is caused by external harmonic forces in $y$-direction (80 pN/atom) with the period of $T=220$~fs in (a) and $T=190$~fs in (b) applied to carbon atoms in the vicinity of $x=0$ (red atoms in the snapshots to the right). In the positive radiation pressure effect, (a), the kink is repelled from the radiation source (see also the top snapshots of molecular dynamics simulations). In the negative radiation pressure effect, (b), the kink is attracted to the radiation source (see also the bottom snapshots of molecular dynamics simulations).
}
\label{fig2}
\end{figure}

Fig. \ref{fig2} (a) and (b) demonstrates two representative results of our calculations exemplifying the positive and negative radiation pressure effects. Both plots were obtained using the same set of parameters except for the driving force period. These plots indicate (by color) $z$-coordinates of the central chain of nanoribbon atoms as functions of time. One can notice that
at the selected force amplitude, $T=220$~fs excitation causes the positive radiation pressure (PRP) effect displacing the kink away from the radiation source (Fig. \ref{fig2}(a)). In the case of a slightly higher frequency ($T=190$~fs, Fig. \ref{fig2}(b)), the kink is attracted to the radiation source exhibiting the negative radiation pressure effect~\cite{Tomasz08a}.

\newpage

\begin{figure*}[t]
\begin{center}
\includegraphics[height=0.3\textwidth]{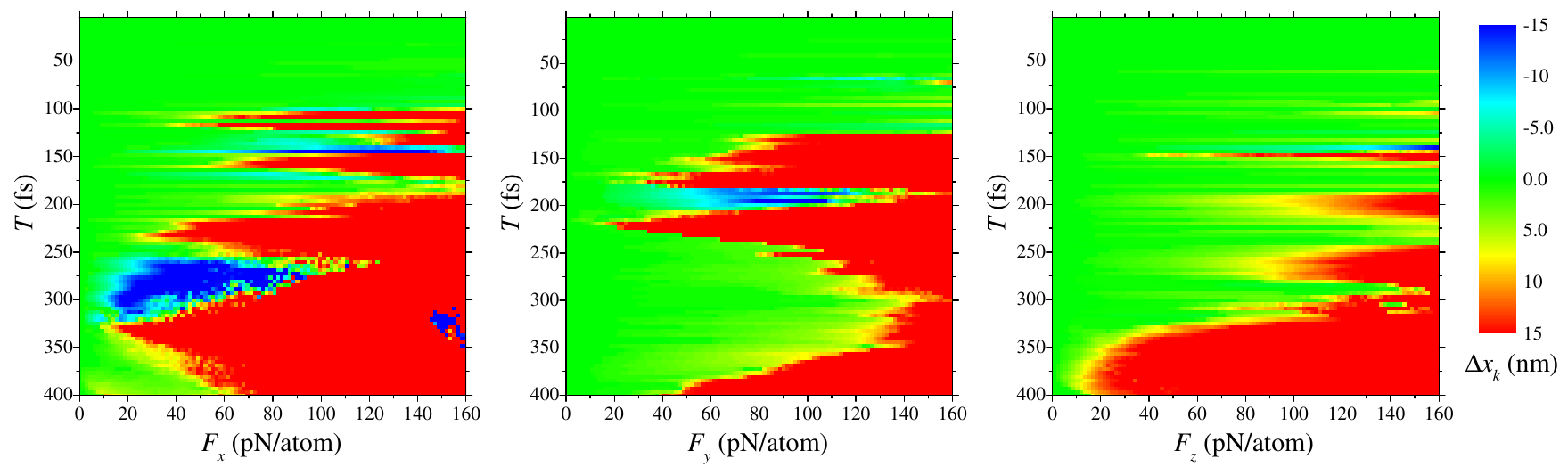}\\
\vspace{2mm}
\includegraphics[width=0.95\textwidth]{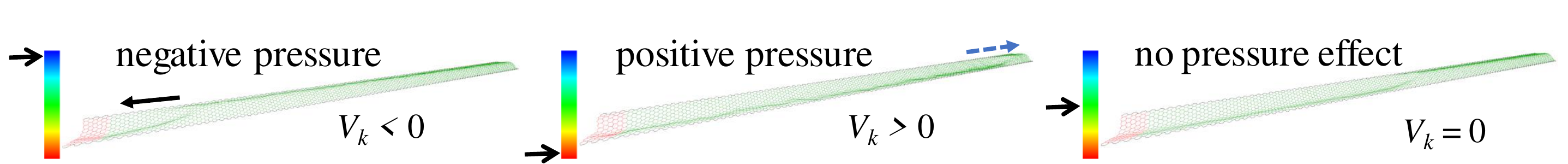}
\end{center}
\caption{The final displacement of the kink, $\Delta x_k$, at $t_f=30$ ps as a function of the applied force amplitude and period. Different plots correspond to different directions of the applied force. The positive  and negative radiation pressure effects are represented by red and blue colors, respectively. Bottom: snapshots of molecular dynamics exemplifying the red, blue and green colors on the image plots above. Here, $V_k$ is the kink velocity. }
\label{fig3}
\end{figure*}

In order to get a systematic understanding of the conditions causing the positive and negative radiation pressure effects, we have performed a series of extensive numerical simulations scanning the force amplitude-force period parameter space. The results of these simulations can be found in Fig. \ref{fig3} showing the final kink displacement. We note that generally, the effect type has a complex dependence on the driving force parameters. Interestingly, especially when excited by $F_y$ and $F_z$ forces, the negative radiation pressure effect can be observed in very narrow intervals of periods. Moreover, the noisy regions in Fig. \ref{fig3} (seen at stronger forces and longer periods) are likely related to resonant oscillations of nanoribbon that strongly interfere with the radiation pressure effects.

\begin{figure}[t]%
(a) \includegraphics[width=0.44\textwidth]{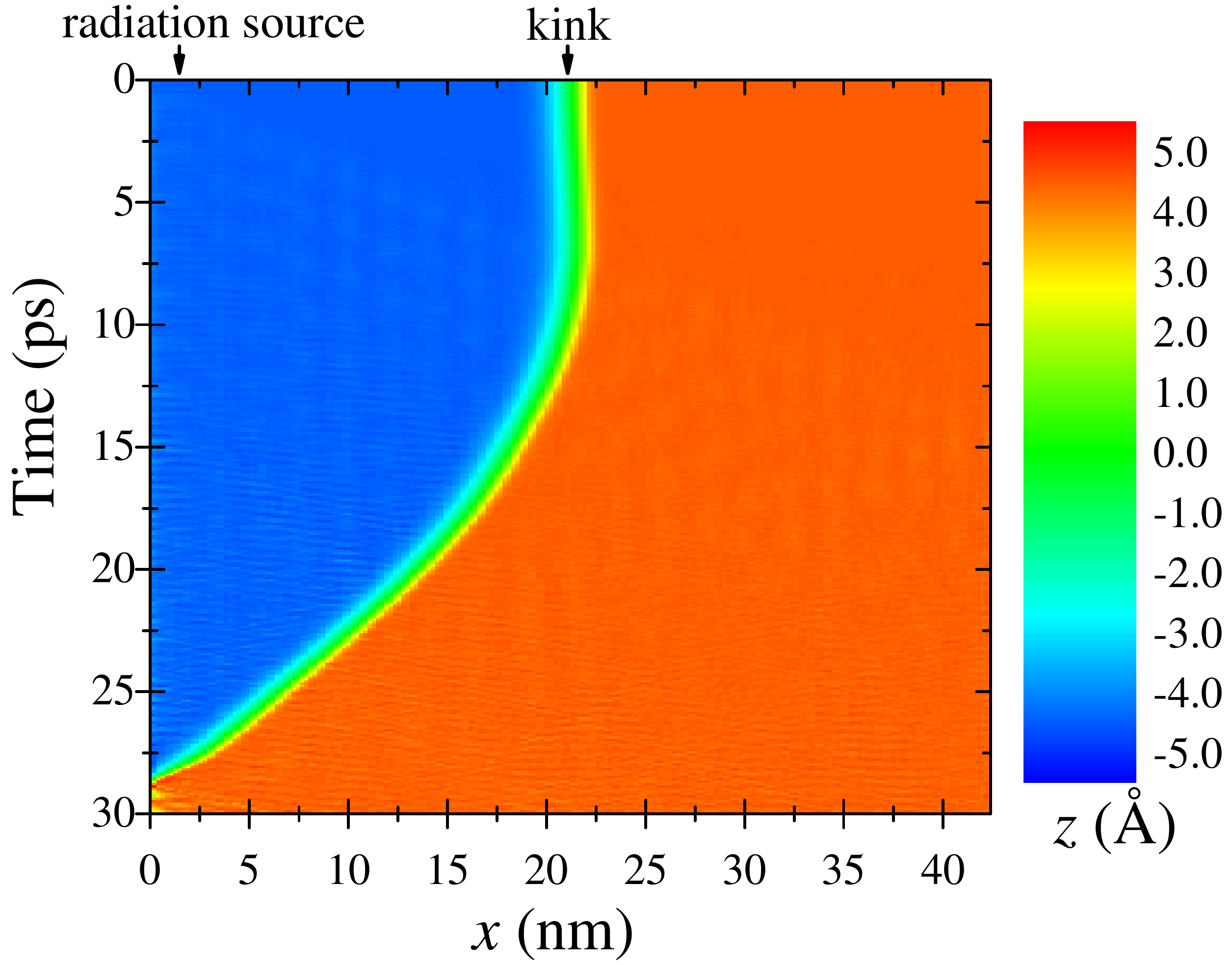}
(b)  \includegraphics[width=0.44\textwidth]{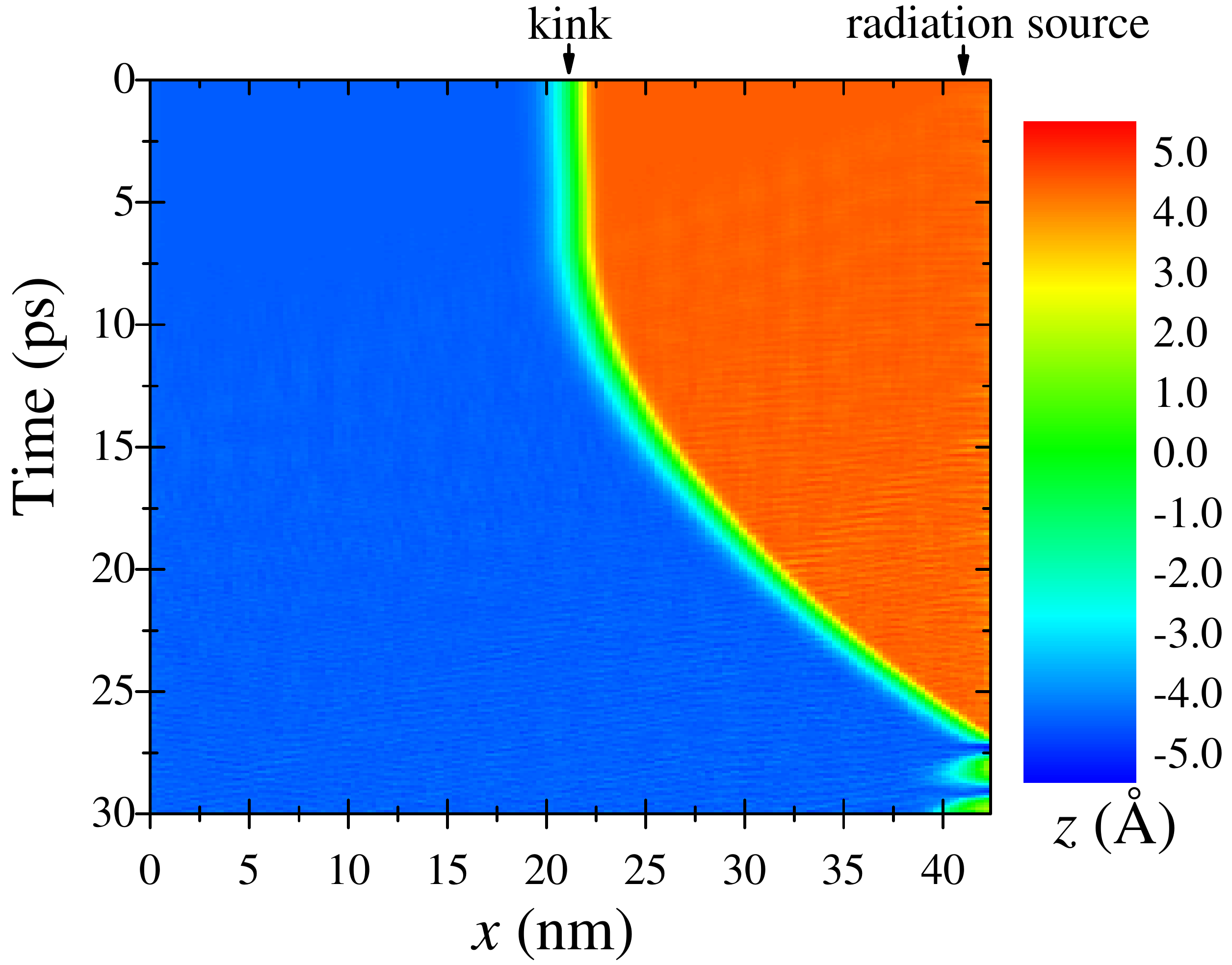}
\caption{NRP effect in the case of radiation incoming from the left edge (a) and right edge (b).  The radiation is caused by external harmonic forces in $y$-direction ($F_y=80$ pN/atom, $T=200$ fs) applied to all carbon atoms within $23$ \r{A} from the left (a) or right (b) edge. Similarly to Fig. \ref{fig2}, the color indicates
$z$-coordinates of the central chain of atoms along the membrane. In both cases the kink (green region) displaces towards the radiation source.}
\label{fig4}
\end{figure}

Additionally, we investigated the symmetry of NRP by changing the location of the radiation source. For this purpose, we used the shorter nanoribbon ($L=42$ \r{A}) with an initially immobile kink located in the middle. The radiation was created near the left edge in one run and near the right edge -- in another.
The results of these simulations are presented in Fig. \ref{fig4}. From Fig. \ref{fig4}, one can observe that the kink is symmetrically attracted to the radiation source.
We conclude that the kink asymmetry in $x$-direction
(see Ref. \cite{Yamaletdinov17b} for more details)  does not influence the manifestation of NRP.
This also excludes the mechanism discussed in \cite{Romanczukiewicz:2017hdu} relevant to the behavior of asymmetric kinks in $\phi^6$ model.

Finally, we have also performed simulations for a  square driving force. In such a case we have also observed NRP proving that the effect is not limited to purely harmonic excitations.

\section{Discussion}
In order to understand the positive and negative radiation pressure effects, one can assume that the origin of these effects is related to the kink deformation by incoming waves. In this scenario,
different deformation types displace kinks in different directions.
Moreover, a kink subjected to the superposition of waves experiences their averaged action.

Consider a kink irradiated by an incoming wave with of a wavelength $\lambda_{in}$. Due to a non-linear interaction with kink, the
transmitted wave contains the incoming frequency and its harmonics: $\lambda_{tr,1}=\lambda_{in}$, $\lambda_{tr,2}=\lambda_{in}/2$, etc.
One can expect the highest harmonics generation rate is achieved when the kink length, $L$, is in  resonance with the wavelength of incoming
radiation, namely, $L=n\lambda_{in}$. Intuitively, among all possible values of $n$, the strongest effect corresponds to $n=1$.

The same picture can be presented from a different point of view. Let us now think that the kink absorbs all incoming radiation from one side and
emits it from the other (and also partially emits it back). These processes are related through an effective phase difference $2 \pi L /\lambda_{in}$.
When $L=n\lambda_{in}$, the 'pressures' of
$\lambda_{in}$ and $\lambda_{tr,1}$ waves on the kink significantly compensate each other. Then, the 'force' associated with  $\lambda_{tr,2}$ pushes
the kink towards the radiation source (the NRP effect).

\begin{figure}[t]%
	\centering \includegraphics[width=0.65\textwidth]{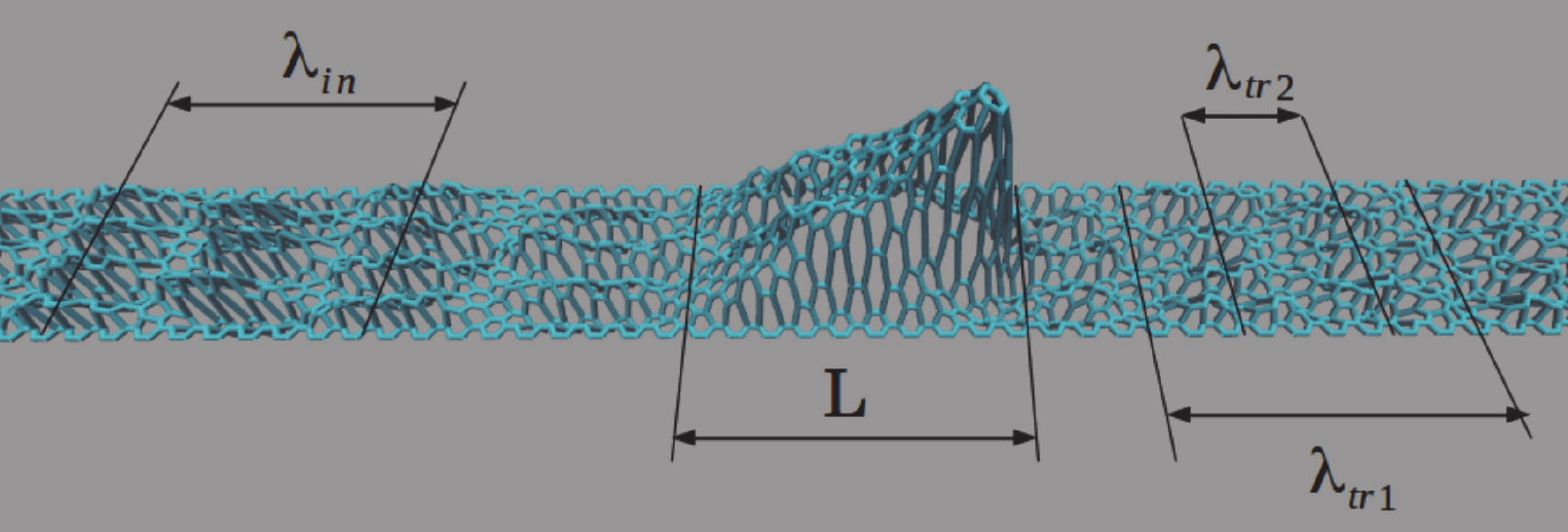}
	\caption{NRP effect in the case of radiation incoming from the left ($T=144$ fs, $F_z=120$ pN/atom). The incoming radiation wavelength
$\lambda_{in}$, kink`s length $L$, and two major modes of transmitted radiation $\lambda_{tr,1} = 2 \lambda_{tr,2}$ are labeled. For illustrative purposes, the $z$-axis is scaled by a factor of 30 compared to $x$ and $y$.}
	\label{fig5}
\end{figure}

In order to resolve the waves propagating through the kink, we plot a scaled difference of atomic positions,
which for $i$-th atom in $n$-th simulation step are taken as $[x_{n,i}; y_{n,i}; z_{n,i}]=[x_{n,i}; y_{n,i}; 30(z_{n,i}-z_{0,i})]$. This quantity is shown in
Fig. \ref{fig5} in the regime of NRP induced by a $z$-direction force
($T=144$~fs, $F_z=120$~pN/atom). One can notice that the transmitted wave contains two major components:
 $\lambda_{tr,1} =42.6$ \r{A} and $\lambda_{tr,2} = \lambda_{tr,1}/2$.  Moreover, the discussed above resonance condition and second harmonic generation can be clearly
distinguished. From these, some important characteristics can be extracted (the wavelength, propagation speed $c_z \approx 3$~km/s, etc.).

Similar results were obtained for $y$-direction force ($T=194$ fs, $F_z=74$ pN/atom), including a similar interaction with $y$-oriented waves. The NRP was induced by $\lambda_{in} =42.6$ \r{A}  wave propagating at $c_y \approx 2.1$~km/s.

We have also analyzed the final position of the kink as a function of the driving force amplitude. In some regions, the NRP is observed at very small amplitudes. However, more frequently, PRP at smaller amplitudes is followed by NRP at intermediate amplitudes transforming back to PRP at higher amplitudes.
For a certain range of  driving force it was possible to fit the final kink's position with $x_f(F)=aF^2-bF^4$, where $F$ is the driving force amplitude. This relation is very similar to the one given by Eq. (\ref{F_phi4}).

Another important observation is that the unidirectional driving force causes atomic oscillations (after the scattering) in all three directions. This means that the scattering on the kink involves several channels indeed and a wave with initially single polarization excites other polarizations as well. In the power spectrum the frequency of the incident wave dominates in all directions. Higher harmonics were also clearly visible. However, different polarizations propagate with different velocities and can carry different momenta.
Therefore, it is possible that a wave with a smaller momentum, after scattering, would generate a wave with a higher momentum (see also Eq. (\ref{eq:force}) and related description).

We also performed additional simulations revealing some vibrational modes of the kink. In these simulations, the kink was perturbed from its static geometry and the oscillations of atoms at its center, in the vicinity and far away from the kink were analyzed. The longitudinal compression (in the $x-$direction) has revealed that the kink oscillates at a frequency \mbox{$f_1=0.96$ THz} ($T=1.04$ ps). Moreover, the power spectrum of oscillations involves another peak at
\mbox{$f_2=2.76$ THz} ($T_2=362$ fs), which is very close to $3f_1$. Two frequencies are associated with  compression in $y-$direction: $f_3=2.26$ THz ($T_3=442$ fs) and $f_4=0.83$ THz ($T_4=1.2$ ps). The sideways-shift deformation leads to oscillations at $f_5=0.33$ THz ($T_5=3.0$ ps). We expect that the waves scattering on the kink can  excite these modes and that the resonances between the modes and the incoming wave are responsible for the regions of the NRP and PRP observed in the Fig. \ref{fig3}. In particular, the region of NRP in the Fig. \ref{fig3} (b) is very just below the frequency of $2f_2$. However the origin of other resonances is not so obvious, especially given the large number of vibrational modes of the kink.

\section{Final remarks}
It follows from our MD simulations and analysis that the graphene kinks are much more complex objected compared to the standard $\phi^4$  kinks. Nonetheless, there are many similarities in their dynamics including NRP and PRP effects. We have outlined two mechanisms, which could be responsible for the NRP:
One is the scattering from  nearly reflectionless kink and the other is the multichannel scattering. The overall picture, however, is also highly disturbed by the presence of vibrational modes, which can become excited at resonances.

In a broader context, small waves and an interplay between the positive and negative radiation pressure effects may have a huge impact on the system of topological defects leading to a fast collapse of an initially almost static  kink-antikink configuration via a chain of annihilation events induced by radiation generated during earlier annihilations~\cite{Romanczukiewicz:2017hdu}.
In higher dimensions, some vacua can be destabilized by the residual radiation leaving the one with the smallest mass parameter as the most stable and favorable final state of evolution (this may have important implications also in a cosmological context).

From the experimental point of view, vibrations of graphene nanoribbon could be generated by applying an AC bias between the nanoribbon and suitably located electrodes. For the graphene nanoribbons  of small sizes (such as these considered in this study), the characteristic amplitudes and frequencies, in principle, are within the limits of modern terahertz electronics~\cite{Dhillon17a}. However, the use of larger membranes (e.g., based on double-layer graphene) would allow for scaling the frequencies down. The buckled graphene membranes with the appropriate geometry could be fabricated using the negative thermal expansion of graphene (opposite to most
materials, graphene contracts on heating and expands on cooling)~\cite{yoon2011negative}. For instance, a thermally oxidized
silicon wafer with an array of lithographically defined U-shaped grooves that are formed by chemical or
plasma etching can be used as substrate. Graphene transferred to the wafer surface at a high temperature will cool,
expand, and buckle above the grooves. Compared to other 2D materials/membranes, the advantages of using graphene include its semi-metallic state enabling electro-mechanical control of its shape, its single-atom thickness providing a strong radiation-kink coupling, and recent advancements in the graphene technology making the graphene the most suitable for such kind of experiments.

Finally, we emphasize that the NRP effect in buckled graphene is fundamentally different from that in small particle-like objects exposed to planar or some conical waves \cite{Ruffner, Gorlach}. Such objects can bend the light (or other type of waves) such that the longitudinal component of the wave increases by the cost of its transversal component. Under certain circumstances particles causing such bending potential experience some NRP-like behavior.

\section*{Acknowledgements}
R.D.Y. thanks the Federal Agency for Scientific Organizations (FASO Russia) for funding.


\section*{References}

\bibliographystyle{elsarticle-num}
\bibliography{memcap}

\end{document}